\documentstyle[graphicx,ulem,amsmath]{mn2e}

\begin{document}

\title[WiggleZ Survey: expansion history]{The WiggleZ Dark Energy
  Survey: Joint measurements of the expansion and growth history at $z
  < 1$}

\author[Blake et al.]{\parbox[t]{\textwidth}{Chris
    Blake$^1$\footnotemark, Sarah Brough$^2$, Matthew Colless$^2$,
    Carlos Contreras$^1$, \\ Warrick Couch$^1$, Scott Croom$^3$,
    Darren Croton$^1$, Tamara M.\ Davis$^4$, \\ Michael
    J.\ Drinkwater$^4$, Karl Forster$^5$, David Gilbank$^6$, Mike
    Gladders$^7$, \\ Karl Glazebrook$^1$, Ben Jelliffe$^3$, Russell
    J.\ Jurek$^8$, I-hui Li$^1$, Barry Madore$^9$, \\ D.\ Christopher
    Martin$^5$, Kevin Pimbblet$^{10}$, Gregory B.\ Poole$^1$, Michael
    Pracy$^3$, \\ Rob Sharp$^{2,11}$, Emily Wisnioski$^1$, David
    Woods$^{12}$, Ted K.\ Wyder$^5$ and H.K.C. Yee$^{13}$} \\ \\ $^1$
  Centre for Astrophysics \& Supercomputing, Swinburne University of
  Technology, P.O. Box 218, Hawthorn, VIC 3122, Australia \\ $^2$
  Australian Astronomical Observatory, P.O. Box 296, Epping, NSW 1710,
  Australia \\ $^3$ Sydney Institute for Astronomy, School of Physics,
  University of Sydney, NSW 2006, Australia \\ $^4$ School of
  Mathematics and Physics, University of Queensland, Brisbane, QLD
  4072, Australia \\ $^5$ California Institute of Technology, MC
  278-17, 1200 East California Boulevard, Pasadena, CA 91125, United
  States \\ $^6$ South African Astronomical Observatory, P.O. Box 9,
  Observatory, 7935, South Africa \\ $^7$ Department of Astronomy and
  Astrophysics, University of Chicago, 5640 South Ellis Avenue,
  Chicago, IL 60637, United States \\ $^8$ Australia Telescope
  National Facility, CSIRO, Epping, NSW 1710, Australia \\ $^9$
  Observatories of the Carnegie Institute of Washington, 813 Santa
  Barbara St., Pasadena, CA 91101, United States \\ $^{10}$ School of
  Physics, Monash University, Clayton, VIC 3800, Australia \\ $^{11}$
  Research School of Astronomy \& Astrophysics, Australian National
  University, Weston Creek, ACT 2611, Australia \\ $^{12}$ Department
  of Physics \& Astronomy, University of British Columbia, 6224
  Agricultural Road, Vancouver, BC V6T 1Z1, Canada \\ $^{13}$
  Department of Astronomy and Astrophysics, University of Toronto, 50
  St.\ George Street, Toronto, ON M5S 3H4, Canada}

\maketitle

\begin{abstract}
  We perform a joint determination of the distance-redshift relation
  and cosmic expansion rate at redshifts $z = 0.44$, $0.6$ and $0.73$
  by combining measurements of the baryon acoustic peak and
  Alcock-Paczynski distortion from galaxy clustering in the WiggleZ
  Dark Energy Survey, using a large ensemble of mock catalogues to
  calculate the covariance between the measurements.  We find that
  $D_A(z) = (1205 \pm 114, 1380 \pm 95, 1534 \pm 107)$ Mpc and $H(z) =
  (82.6 \pm 7.8, 87.9 \pm 6.1, 97.3 \pm 7.0)$ km s$^{-1}$ Mpc$^{-1}$
  at these three redshifts.  Further combining our results with other
  baryon acoustic oscillation and distant supernovae datasets, we use
  a Monte Carlo Markov Chain technique to determine the evolution of
  the Hubble parameter $H(z)$ as a stepwise function in 9 redshift
  bins of width $\Delta z = 0.1$, also marginalizing over the spatial
  curvature.  Our measurements of $H(z)$, which have precision better
  than $7\%$ in most redshift bins, are consistent with the expansion
  history predicted by a cosmological-constant dark-energy model, in
  which the expansion rate accelerates at redshift $z < 0.7$.
\end{abstract}

\section{Introduction}
\renewcommand{\thefootnote}{\fnsymbol{footnote}}
\setcounter{footnote}{1}
\footnotetext{E-mail: cblake@astro.swin.edu.au}

One of the fundamental goals of observational cosmology is to
determine the expansion rate of the Universe as a function of
redshift.  Measurements of the expansion history, which can be
described by the evolution of the Hubble parameter $H(z) = (1+z) \,
da/dt$ with redshift $z$, where $a(t)$ is the cosmic scale factor at
time $t$, provide one of the most important observational tests of the
cosmological models which characterize the different components of the
Universe and their evolution with time.  In particular, a paramount
problem in cosmology is to understand the physical significance of the
``dark energy'' which appears to dominate the cosmic energy density
today, as described by the phenomenology of the standard
cosmological-constant cold-dark-matter ($\Lambda$CDM) model.

A number of powerful tools to measure the cosmic expansion history
beyond the local Universe have been developed in recent decades.
Foremost amongst these probes is the use of distant Type Ia supernovae
(SNe) as standard candles (e.g.\ Riess et al.\ 1998, Perlmutter et
al.\ 1999, Amanullah et al.\ 2010, Conley et al.\ 2011, Suzuki et
al.\ 2011).  The apparent peak magnitude of these supernovae,
following certain corrections based on the light-curve shape which
decrease the observed scatter in the peak brightness, yield a relative
luminosity distance as a function of redshift, i.e. $D_L(z) \, H_0/c$
where $D_L(z)$ is the luminosity distance, $H_0$ is the local Hubble
parameter and $c$ is the speed of light.

Although such measurements accurately trace the shape of the
distance-redshift ``Hubble diagram'' in the redshift range $z < 1$, a
number of qualifications must be mentioned.  Firstly, the expansion
history $H(z)$ is not directly measured by supernovae but must be
determined as a derivative of the noisy luminosity distances (Wang \&
Tegmark 2005, Sollerman et al.\ 2009, Shafieloo \& Clarkson 2010).
Secondly, obtaining the expansion rate from the luminosity distance
requires an additional assumption about spatial curvature, which
influences the geodesics followed by photons.  Thirdly, despite the
impressive and thorough treatment in recent supernovae analyses of the
systematic errors which could bias cosmological fits, these
systematics now limit the utility of these datasets.

Large galaxy surveys offer a complementary route for mapping cosmic
distances and expansion, using two principal techniques.  Firstly, the
large-scale clustering pattern of galaxies contains the signature of
baryon acoustic oscillations (BAOs), a preferred length scale
imprinted in the distribution of photons and baryons by the
propagation of sound waves in the relativistic plasma of the early
Universe.  This length scale, the sound horizon at the baryon drag
epoch $r_s(z_d)$, may be accurately calibrated by observations of the
Cosmic Microwave Background (CMB) radiation and applied as a
cosmological standard ruler (Blake \& Glazebrook 2003, Seo \&
Eisenstein 2003).  Some applications to galaxy datasets are presented
by Eisenstein et al.\ (2005), Percival et al.\ (2010), Beutler et
al.\ (2011), Blake et al.\ (2011b), Padmanabhan et al.\ (2012) and
Anderson et al.\ (2012).

Given a sufficiently-large galaxy survey at a redshift $z$, the
preferred scale may be detected in both the tangential direction on
the sky as an enhancement in the number of galaxy pairs with a given
angular separation $\Delta \theta$, and in the radial direction as an
excess of pairs with redshift separation $\Delta z$.  If these two
signals can be simultaneously extracted by measuring galaxy clustering
in tangential and radial bins, they respectively carry information
about the angular diameter distance $D_A(z) = D_L(z)/(1+z)^2$ and the
Hubble expansion rate at the survey redshift in units of the standard
ruler, $r_s(z_d)/[(1+z)D_A(z)] \sim \Delta \theta$ and $r_s(z_d) H(z)
\sim c \Delta z$.  If the galaxy survey only permits the baryon
acoustic peak to be detected in the angle-averaged galaxy clustering
pattern, then an effective ``dilation scale'' distance is measured
which consists of two parts $D_A(z)$ and one part $1/H(z)$: $D_V(z) =
[ (1+z)^2 D_A(z)^2 cz/H(z) ]^{1/3}$ (Eisenstein et al.\ 2005,
Padmanabhan \& White 2008).

The second technique through which large-scale structure surveys
permit measurement of geometrical distances is the ``Alcock-Paczynski
(AP) test'' (Alcock \& Paczynski 1979).  The AP test probes the
cosmological model by comparing the observed tangential and radial
dimensions of objects which are assumed to be isotropic in the correct
choice of model.  It can be applied to the 2-point statistics of
galaxy clustering if redshift-space distortions, the principal
additional source of anisotropy, can be successfully modelled
(Ballinger, Peacock \& Heavens 1996, Matsubara \& Suto 1996, Simpson
\& Peacock 2010).  By equating radial and tangential physical scales,
independently of any underlying standard ruler, the observable $\Delta
z/\Delta \theta \sim (1+z) D_A(z) H(z)/c$ may be determined (Outram et
al.\ 2004, Marinoni \& Buzzi 2010, Blake et al.\ 2011c).

Therefore, using a combination of BAO and/or AP measurements,
large-scale galaxy surveys can supply independent measurements of the
distance-redshift relation $D_A(z)$ and expansion history $H(z)$.  We
note that the Hubble expansion rate as a function of redshift may also
be inferred from the relative ages of passively-evolving galaxies
(Jimenez \& Loeb 2002, Stern et al.\ 2010, Carson \& Nichol 2010,
Moresco et al.\ 2012).  This is a promising technique albeit subject
to assumptions about the stellar populations of these galaxies, in
particular about galaxy metallicity at high redshift.

We focus here on distance measurements using the WiggleZ Dark Energy
Survey (Drinkwater et al.\ 2010), which was designed to extend the
study of large-scale structure over large cosmic volumes to redshifts
$z > 0.5$.  The study presented here builds upon two existing
distance-scale measurements using the WiggleZ dataset: Blake et
al.\ (2011b) reported the measurement of the angle-averaged baryon
acoustic peak at redshifts $z = (0.44, 0.6, 0.73)$, and Blake et
al.\ (2011c) applied the Alcock-Paczynski test to the 2D clustering
power spectrum.  In this latter study we combined the AP fits with SNe
data to estimate the Hubble parameter relative to its local value,
$H(z)/H_0$.  We here extend these analyses by combining the WiggleZ
measurements of $D_A^2/H$ and $D_A H$, including a calculation of the
covariance of the statistics, to extract measurements of $D_A(z)$ and
$H(z)$ as a function of redshift, in absolute units independent of the
value of $H_0$, based solely on WiggleZ Survey data (and a
sound-horizon calibration).  Furthermore, by combining these
measurements with SNe and other galaxy datasets, we constrain the
cosmic expansion rate $H(z)$ as a stepwise function in the redshift
range $z < 0.9$, and fit a variety of cosmological models to the
results.

\section{Data}
\label{secdata}

\begin{figure*}
\begin{center}
\resizebox{16cm}{!}{\rotatebox{270}{\includegraphics{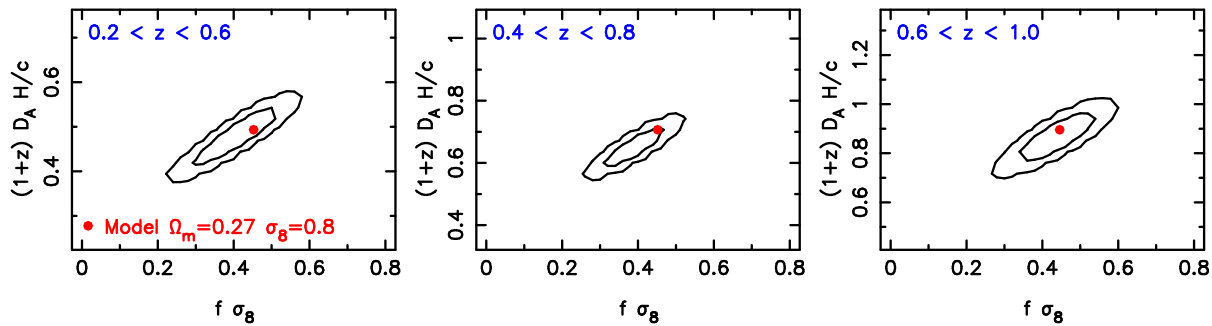}}}
\end{center}
\caption{The joint likelihood of the Alcock-Paczynski scale distortion
  parameter $F(z) \equiv (1+z) D_A(z) H(z)/c$ and the normalized
  growth rate quantified by $f \, \sigma_8(z)$, obtained from fits to
  the 2D galaxy power spectra of the WiggleZ Dark Energy Survey in
  three overlapping redshift slices $0.2 < z < 0.6$, $0.4 < z < 0.8$
  and $0.6 < z < 1.0$.  This Figure was produced by marginalizing over
  the linear bias factor $b^2$.  The probability density is plotted as
  contours enclosing $68.27\%$ and $95.45\%$ of the total likelihood.
  The solid circles indicate the parameter values in a fiducial flat
  $\Lambda$CDM cosmological model with parameters $\Omega_{\rm m} =
  0.27$, $\sigma_8 = 0.8$.}
\label{figap}
\end{figure*}

\subsection{WiggleZ Survey clustering measurements}

The WiggleZ Dark Energy Survey (Drinkwater et al.\ 2010) is a
large-scale redshift survey of bright emission-line galaxies which was
carried out at the Anglo-Australian Telescope between August 2006 and
January 2011.  The galaxy sample utilized by this study is drawn from
the final set of observations covering about 800 deg$^2$ of sky in six
regions, including a total of $N = 158{,}741$ galaxies in the redshift
range $0.2 < z < 1.0$, and is the same dataset as used for the
analysis of the baryon acoustic peak by Blake et al.\ (2011b).  Our
study is based on measurements of the angle-averaged galaxy
correlation function and 2D galaxy power spectrum in tangential and
radial Fourier bins in three overlapping redshift ranges $0.2 < z <
0.6$, $0.4 < z < 0.8$ and $0.6 < z < 1.0$.  The effective redshifts of
the measurements in these three redshift slices are $z_{\rm eff} =
(0.44, 0.6, 0.73)$ (Blake et al.\ 2011b).  We use overlapping, wide
redshift ranges in order to ensure a detection of the baryon acoustic
peak in each redshift slice (following Percival et al.\ 2010) and to
provide the best mapping of the distance-redshift relation.  We
account for the correlations between the measurements when fitting
models.

\subsection{Fitting the baryon acoustic peak}

Blake et al.\ (2011b) presented our analysis of the WiggleZ galaxy
correlation function to map the baryon acoustic peak in these three
redshift ranges.  The correlation functions contain evidence for the
baryon acoustic peak in each redshift slice.  Our model for fitting
the correlation function measurements to determine the standard-ruler
distance is described in Section 3.1 of Blake et al.\ (2011b).  Our
results are most cleanly expressed as a measurement of the acoustic
parameter,
\begin{equation}
A(z) = \frac{100 \, D_V(z) \sqrt{\Omega_{\rm m} h^2}}{c\, z} ,
\label{eqaz}
\end{equation}
at the effective redshift of the sample.  We marginalized over the
shape of the clustering pattern (parameterized by the physical matter
density $\Omega_{\rm m} h^2$), non-linear damping of the acoustic
peak, and galaxy bias.  The results for $A(z)$ are listed in Table
\ref{tabres}, reproduced from Blake et al.\ (2011b).

\begin{table*}
\begin{center}
\caption{Results of cosmological model fits to the galaxy correlation
  functions and power spectra measured in three overlapping redshift
  slices of the WiggleZ Dark Energy Survey.  The acoustic scale
  parameter $A(z)$ is obtained from the fit to the baryon acoustic
  peak in the correlation function (marginalizing over the matter
  density $\Omega_{\rm m} h^2$, the damping parameter $\sigma_v$ and
  the galaxy bias factor) and the parameters $F(z)$ and $f \,
  \sigma_8(z)$ are measured using the 2D power spectra (marginalizing
  over a galaxy bias which is not assumed to be identical to its value
  in the correlation function fit).  The angular diameter distance
  $D_A(z)$ and Hubble expansion rate $H(z)$ are derived from the
  measurements of $A(z)$ and $F(z)$ assuming a CMB-motivated prior in
  $\Omega_{\rm m} h^2$ in order to calibrate the standard ruler.}
\label{tabres}
\begin{tabular}{cccc}
\hline
Redshift slice & $0.2 < z < 0.6$ & $0.4 < z < 0.8$ & $0.6 < z < 1.0$ \\
\hline
Effective redshift $z$ & $0.44$ & $0.60$ & $0.73$ \\
$A(z) \equiv 100 D_V(z)\sqrt{\Omega_{\rm m} h^2}/cz$ & $0.474 \pm 0.034$ & $0.442 \pm 0.020$ & $0.424 \pm 0.021$ \\
$F(z) \equiv (1+z)D_A(z)H(z)/c$ & $0.482 \pm 0.049$ & $0.650 \pm 0.053$ & $0.865 \pm 0.073$ \\
$f \, \sigma_8(z)$ & $0.413 \pm 0.080$ & $0.390 \pm 0.063$ & $0.437 \pm 0.072$ \\
$D_A(z)$ [Mpc] & $1204.9 \pm 113.6$ & $1380.1 \pm 94.8$ & $1533.7 \pm 106.8$ \\
$H(z)$ [km s$^{-1}$ Mpc$^{-1}$] & $82.6 \pm 7.8$ & $87.9 \pm 6.1$ & $97.3 \pm 7.0$ \\
\hline
\end{tabular}
\end{center}
\end{table*}

\subsection{Fitting the 2D power spectrum}

Blake et al.\ (2011c) presented an analysis of the 2D WiggleZ galaxy
power spectrum, where modes are binned by Fourier wavenumber and angle
to the line-of-sight.  We repeated these measurements for the new
choice of redshift bins consistent with the correlation function
analysis.  Our model for fitting the 2D power spectrum to extract the
Alcock-Paczynski distortion is described in Section 3.2 of Blake et
al.\ (2011c).  At each redshift we obtain a measurement of the
distortion parameter $F(z) = (1+z) D_A(z) H(z)/c$ and the normalized
growth rate $f \, \sigma_8$, which quantifies the amplitude of
redshift-space distortions in terms of the growth rate $f$ and
amplitude of matter fluctuations $\sigma_8$.  We also marginalized
over a linear bias factor which we do not require to be identical to
the bias in the correlation function model.  This is a conservative
choice which reflects the possibility that the amplitude of these two
statistics due to galaxy bias and redshift-space distortions may be
scale-dependent.  Figure \ref{figap} displays the joint likelihoods of
$F$ and $f \, \sigma_8$ fitted in each of the new redshift slices.  We
note that although there is a strong correlation between the
parameters, both may be successfully determined, and the results are
collected in Table \ref{tabres}.

\section{Joint fits for the expansion and growth history}

\subsection{Covariances between fitted parameters}

Given that our measurements of the baryon acoustic scale and
tangential/radial clustering anisotropy have taken place in
overlapping redshift slices, using clustering statistics which are
potentially correlated by common cosmic variance, it is important to
determine the covariances between the measurements of $(A, F,
f\sigma_8)$ within and between redshift slices.  This is achieved by
repeating the different parameter fits for each of a series of
lognormal realizations, and using the statistical ensemble to
determine the various correlation coefficients.  We generated 400
lognormal realizations for each WiggleZ survey region and redshift
slice (i.e., 2400 realizations for each redshift slice, or 7200 in
total) using the methods described by Blake et al.\ (2011a).
Lognormal realizations provide a reasonably accurate galaxy clustering
model for the linear and quasi-linear scales which are important for
modelling the large-scale clustering pattern.

\begin{figure*}
\begin{center}
\resizebox{16cm}{!}{\rotatebox{270}{\includegraphics{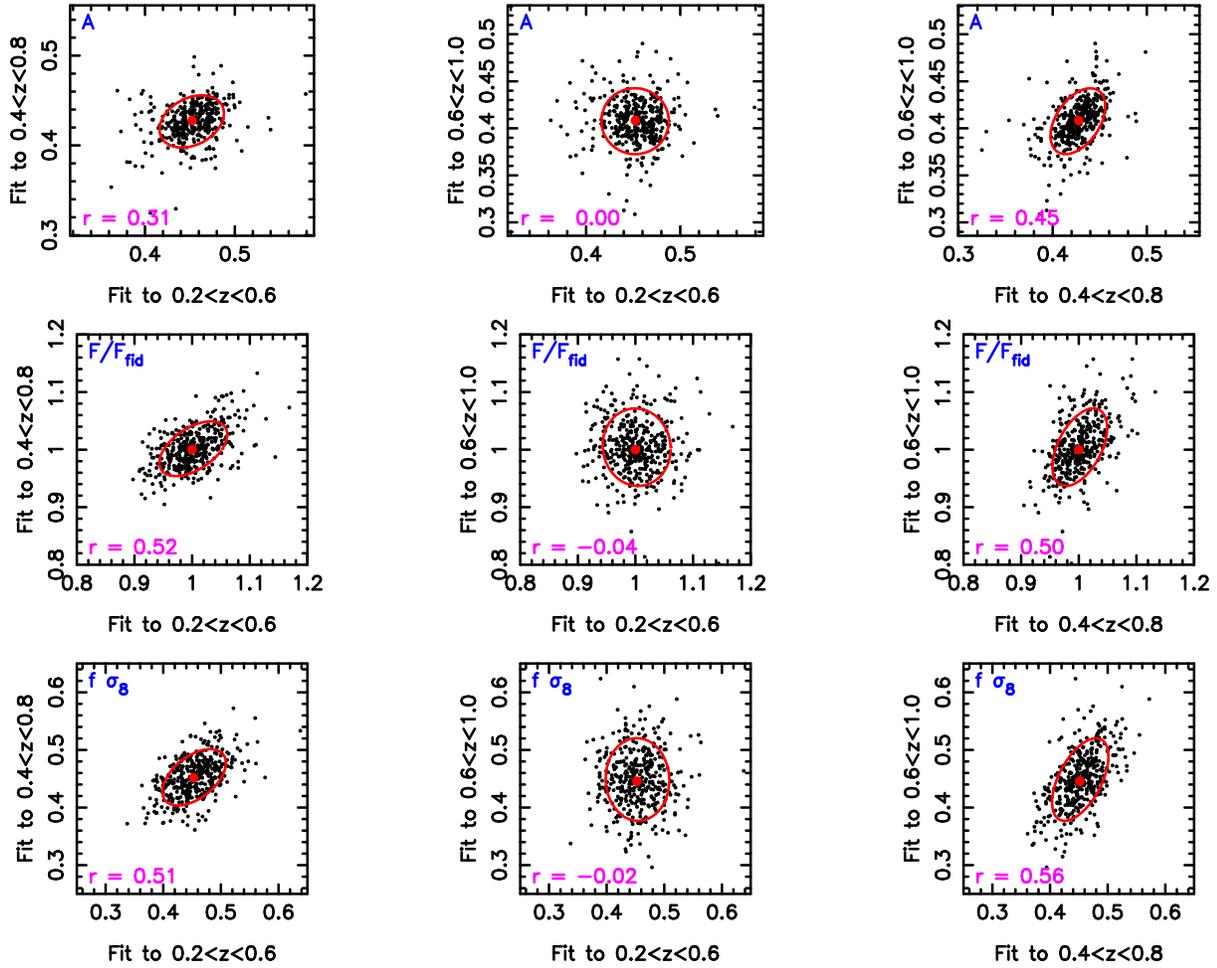}}}
\end{center}
\caption{Correlations of the parameters $A(z)$, $F(z)$ and $f \,
  \sigma_8(z)$ when each parameter is fitted to pairs of the three
  overlapping WiggleZ redshift slices.  The Alcock-Paczynski
  distortion parameter $F$ is plotted relative to its value in the
  fiducial cosmology, $F_{\rm fid}$.  Each small dot represents the
  best-fitting values of the parameters using the correlation
  functions and power spectra measured from 400 independent lognormal
  realizations.  The red ellipses represent the derived covariances
  between the parameter fits, and the solid red circle is the input
  fiducial model of the lognormal realizations.  The correlation
  coefficient $r$ is quoted in the bottom left-hand corner of each
  panel, and is consistent with zero in the second column when
  non-overlapping redshift slices are used.}
\label{figcovpar1}
\end{figure*}

Figure \ref{figcovpar1} displays the correlations between single
parameters fit to different pairs of redshift slices.  In each panel,
the small dots represent the best-fitting pairs of parameter values in
the redshift slices for the 400 lognormal realizations.  The red
ellipses are 2D Gaussians representing the covariance between the
fitted parameters, and the solid red circle is the input fiducial
model used to generate the lognormal realizations.  The correlation
coefficient $r$ is quoted in the bottom left-hand corner of each
panel, and is consistent with zero in the second column when
non-overlapping redshift slices are used.

\begin{figure*}
\begin{center}
\resizebox{16cm}{!}{\rotatebox{270}{\includegraphics{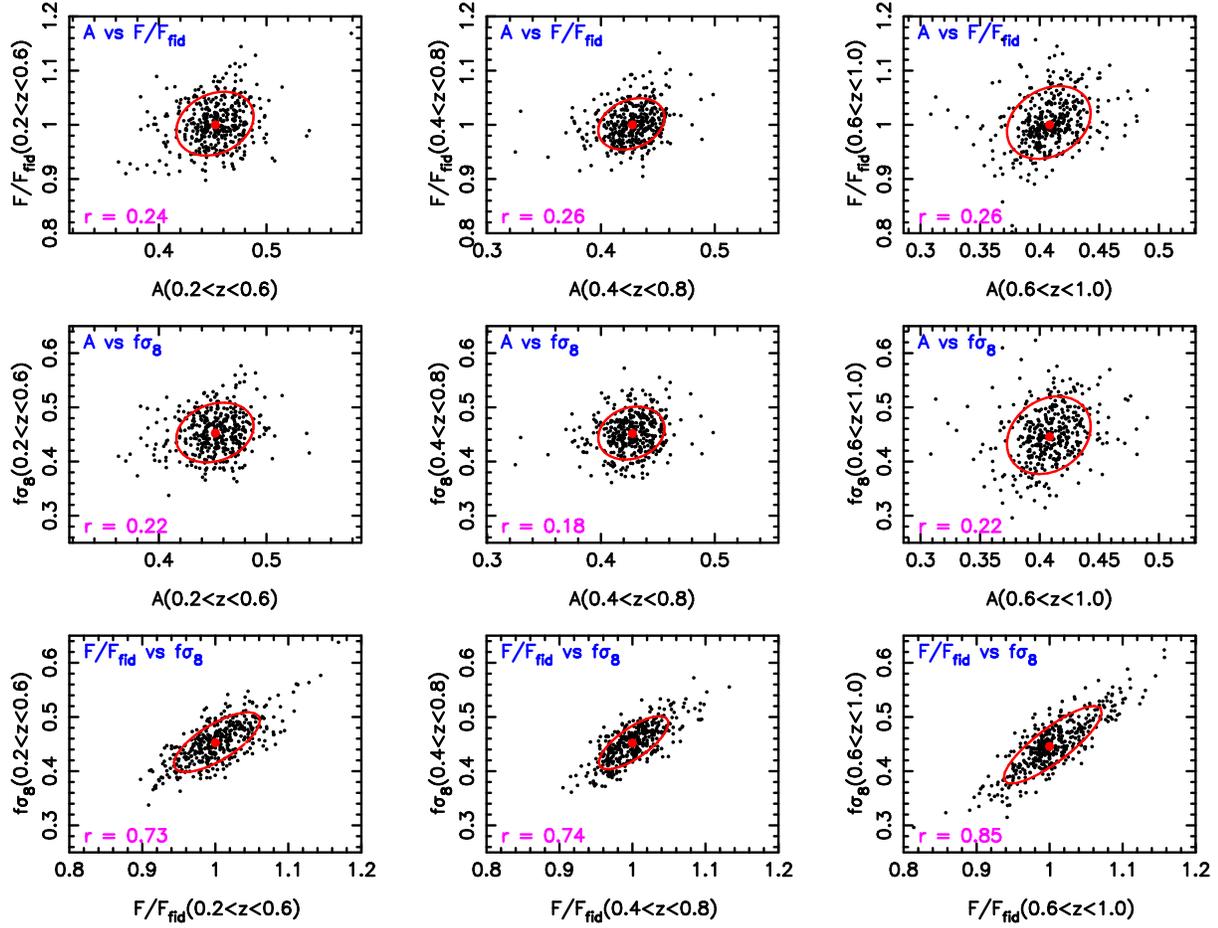}}}
\end{center}
\caption{Correlations between different pairs of the parameters
  $A(z)$, $F(z)$ and $f \, \sigma_8(z)$ fitted to the three WiggleZ
  redshift slices.  The Alcock-Paczynski distortion parameter $F$ is
  plotted relative to its value in the fiducial cosmology, $F_{\rm
    fid}$.  Each small dot represents the best-fitting values of the
  parameters using the correlation functions and power spectra
  measured from 400 independent lognormal realizations.  The red
  ellipses represent the derived covariances between the measurements,
  and the solid red circle is the input fiducial model of the
  lognormal realizations.  The correlation coefficient $r$ is quoted
  in the bottom left-hand corner of each panel.  Although the
  strongest correlation is obtained between $F(z)$ and $f \,
  \sigma_8(z)$, weaker but non-zero correlations are measured between
  both of these parameters and $A(z)$.}
\label{figcovpar2}
\end{figure*}

Figure \ref{figcovpar2} shows the correlations between pairs of
different parameters fit in the same redshift slice, using the same
presentation format as Figure \ref{figcovpar1}.  The strongest
covariance is measured between the Alcock-Paczynski distortion $F$ and
growth rate $f \, \sigma_8$ from redshift-space distortions, with
correlation coefficients $r \sim 0.8$.  The measurements of the baryon
acoustic peak ``monopole'' parameter $A$ are correlated with each of
the ``quadrupole'' parameters $(F,f\sigma_8)$ at a lower level $r \sim
0.2$.  The full $9 \times 9$ covariance matrix for the parameters is
listed in Table \ref{tabcov}.

\begin{figure*}
\begin{center}
\resizebox{16cm}{!}{\rotatebox{270}{\includegraphics{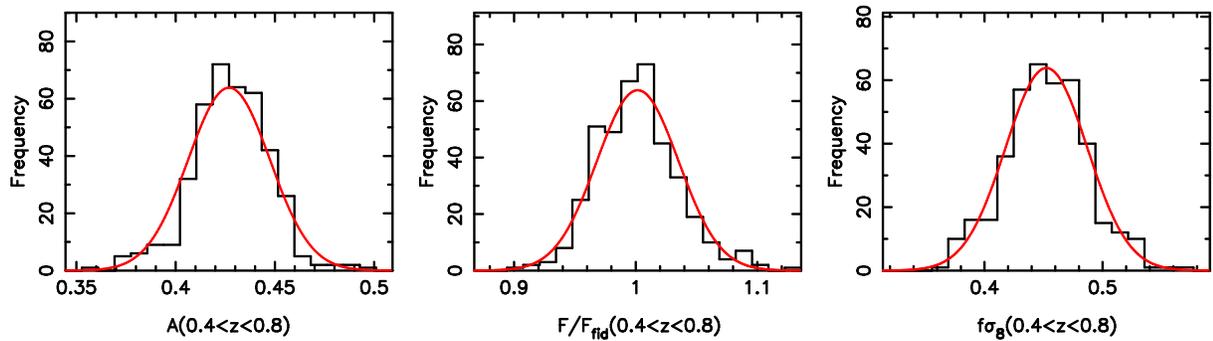}}}
\end{center}
\caption{The 1D distribution of the best-fitting parameters for 400
  lognormal realizations for the second redshift slice, compared to
  the adopted multivariate Gaussian model for the covariance.  The
  full distribution does not contain significant wings that might
  cause confidence regions to be under-estimated in subsequent
  cosmological parameter fits.}
\label{figpardist}
\end{figure*}

Figure \ref{figpardist} plots the 1D distributions of best-fitting
parameters for the second redshift slice, demonstrating that this is
well-described by the multivariate Gaussian model and does not contain
significant wings that might cause the confidence regions to be
under-estimated in subsequent cosmological parameter fits.  The
distributions for the other redshift slices are similar.

\begin{table*}
\begin{center}
\caption{This Table lists the values of $10^3 \uuline{C}$, where
  $\uuline{C}$ is the covariance matrix of measurements from the
  WiggleZ Survey data of the acoustic parameter $A(z)$, the
  Alcock-Paczynski distortion parameter $F(z)$ and normalized growth
  rate $f \sigma_8(z)$, where each parameter is measured in three
  overlapping redshift slices $(z_1,z_2,z_3)$ with effective redshifts
  $z_{\rm eff} = 0.44$, $0.6$ and $0.73$, respectively, where $z_1 =
  [0.2,0.6]$, $z_2 = [0.4,0.8]$ and $z_3 = [0.6,1.0]$.  The data
  vector is ordered such that $\uline{Y}_{\rm obs} = (A_1, A_2, A_3,
  F_1, F_2, F_3, f \sigma_{8,1}, f \sigma_{8,2}, f \sigma_{8,3}) =
  (0.474, 0.442, 0.424, 0.482, 0.650, 0.865, 0.413, 0.390, 0.437)$.
  The chi-squared statistic for any cosmological model vector
  $\uline{Y}_{\rm mod}$ can be obtained via the matrix multiplication
  $\chi^2 = (\uline{Y}_{\rm obs} - \uline{Y}_{\rm mod})^T
  \uuline{C}^{-1} (\uline{Y}_{\rm obs} - \uline{Y}_{\rm mod})$.  The
  matrix is symmetric; we just quote the upper diagonal.}
\label{tabcov}
\begin{tabular}{cccccccccc}
\hline
Parameter & $A(z_1)$ & $A(z_2)$ & $A(z_3)$ & $F(z_1)$ & $F(z_2)$ & $F(z_3)$ & $f\sigma_8(z_1)$ & $f\sigma_8(z_2)$ & $f\sigma_8(z_3)$ \\
\hline
$A(z_1)$         &   1.156 &   0.211 &   0.000 &   0.400 &   0.234 &   0.000 &   0.598 &   0.129 &   0.000 \\
$A(z_2)$         & -       &   0.400 &   0.189 &   0.118 &   0.276 &   0.336 &   0.080 &   0.227 &   0.230 \\
$A(z_3)$         & -       & -       &   0.441 &   0.000 &   0.167 &   0.399 &   0.000 &   0.146 &   0.333 \\
$F(z_1)$         & -       & -       & -       &   2.401 &   1.350 &   0.000 &   2.862 &   1.080 &   0.000 \\
$F(z_2)$         & -       & -       & -       & -       &   2.809 &   1.934 &   1.611 &   2.471 &   1.641 \\
$F(z_3)$         & -       & -       & -       & -       & -       &   5.329 &   0.000 &   1.978 &   4.468 \\
$f\sigma_8(z_1)$ & -       & -       & -       & -       & -       & -       &   6.400 &   2.570 &   0.000 \\
$f\sigma_8(z_2)$ & -       & -       & -       & -       & -       & -       & -       &   3.969 &   2.540 \\
$f\sigma_8(z_3)$ & -       & -       & -       & -       & -       & -       & -       & -       &   5.184 \\
\hline
\end{tabular}
\end{center}
\end{table*}

\subsection{Determination of $D_A(z)$ and $H(z)$}

Using the joint measurements of the Alcock-Paczynski distortion
parameter $F \propto D_A H$ and acoustic parameter $A \propto
(D_A^2/H)^{1/3}$ in each redshift slice, we can break the degeneracy
between the angular-diameter distance $D_A(z)$ and Hubble parameter
$H(z)$.  We fit for $D_A$ and $H$ in each redshift slice using these
measurements and their covariance.  We also marginalize over the
physical matter density $\Omega_{\rm m} h^2$, which appears in
Equation \ref{eqaz} for $A(z)$, using a Gaussian prior with mean
$0.1345$ and width $0.0055$.  This prior is motivated by fits to the
CMB (Komatsu et al.\ 2009) and is independent of the low-redshift
expansion history under certain general assumptions, which are listed
in Section 5.4.1 of Komatsu et al.\ (2009).

\begin{figure}
\begin{center}
\resizebox{8cm}{!}{\rotatebox{270}{\includegraphics{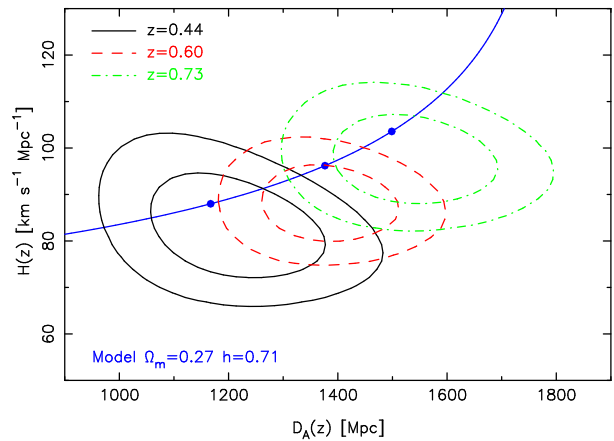}}}
\end{center}
\caption{The joint likelihood of fits of $D_A(z)$ and $H(z)$ to the
  baryon acoustic peak and Alcock-Paczynski distortions in each of
  three overlapping WiggleZ redshift slices.  The two contour levels
  in each case enclose regions containing $68.27\%$ and $95.45\%$ of
  the total likelihood.  A flat $\Lambda$CDM model prediction for
  cosmological parameters $\Omega_{\rm m} = 0.27$ and $h = 0.71$ is
  plotted as the solid line, with circles representing the effective
  redshifts of the three data slices.}
\label{figprobdah}
\end{figure}

The joint likelihood of $D_A(z)$ and $H(z)$ in each of the three
redshift slices is displayed in Figure \ref{figprobdah}, where the
solid line represents the joint variation of these parameters with
redshift in a fiducial cosmological model $\Omega_{\rm m} = 0.27$ and
$h = 0.71$, and the solid circles superimposed on the line indicate
the model prediction for the three analyzed redshift slices.  The
marginalized values of $D_A$ and $H$ at redshifts $z = (0.44, 0.60,
0.73)$ are $D_A(z) = (1205 \pm 114, 1380 \pm 95, 1534 \pm 107)$ Mpc
and $H(z) = (82.6 \pm 7.8, 87.9 \pm 6.1, 97.3 \pm 7.0)$ km s$^{-1}$
Mpc$^{-1}$.  A steady increase in the value of $H(z)$ with $z$ is
consistent with accelerating expansion given that $dH/dt = -H^2 \left[
  1 + (\ddot{a}/a) \right]$ is negative when $\ddot{a} > 0$.  These
measurements of $D_A(z)$ and $H(z)$ are listed in Table \ref{tabres}
along with the marginalized measurements of the normalized growth
rate.  The fractional accuracies with which the parameters are
measured are $7-9\%$ for $D_A$ and $H$, and $16-20\%$ for $f \,
\sigma_8$.  We note that readers wishing to include our dataset in
cosmological parameter fits should use the raw measurements of $(A, F,
f\sigma_8)$ given in Table \ref{tabres}, together with the covariance
matrix listed in Table \ref{tabcov}, rather than these derived values
of $D_A$ and $H$.

\section{Cosmological model fits}
\label{seccosmo}

We now use our joint WiggleZ measurements of the baryon acoustic peak
and Alcock-Paczynski distortions to place constraints on parametric
and non-parametric cosmological models, both alone and in combination
with other datasets.

\subsection{Other cosmological datasets}

Together with the new WiggleZ results described in this study, we add
BAO distance measurements obtained from the 6-degree Field Galaxy
Survey (6dFGS; Beutler et al.\ 2011) and by applying
``reconstruction'' to the sample of Sloan Digital Sky Survey (SDSS)
Luminous Red Galaxies (Padmanabhan et al.\ 2012).  We also include the
joint BAO and AP measurements recently reported by the Baryon
Oscillation Spectroscopic Survey (BOSS, Reid et al.\ 2012).

We additionally use the ``Union 2'' compilation of supernovae data by
Amanullah et al.\ (2010), as obtained from the website {\tt
  http://supernova.lbl.gov/Union}.  This compilation of 557 supernovae
includes data from Hamuy et al.\ (1996), Riess et al.\ (1999, 2007),
Astier et al.\ (2006), Jha et al.\ (2006), Wood-Vasey et al.\ (2007),
Holtzman et al.\ (2008), Hicken et al.\ (2009) and Kessler et
al.\ (2009).  When fitting cosmological models to this SNe dataset we
used the full covariance matrix of these measurements including
systematic errors, as reported by Amanullah et al.\ (2010).  We also
performed an analytic marginalization over the unknown absolute
normalization (Goliath et al.\ 2001, Bridle et al.\ 2002).

Finally, in some fits we include CMB data using the Wilkinson
Microwave Anisotropy Probe (WMAP) ``distance priors'' (Komatsu et
al.\ 2009) using the 7-year WMAP results (Komatsu et al.\ 2011).  The
distance priors quantify the complete CMB likelihood via a 3-parameter
covariance matrix for the acoustic index $\ell_A$, the shift parameter
${\mathcal R}$ and the redshift of recombination $z_*$, as given in
Table 10 of Komatsu et al.\ (2011).  When deriving these quantities we
assumed a physical baryon density $\Omega_{\rm b} h^2 = 0.0226$, a CMB
temperature $T_{\rm CMB} = 2.725 K$ and a number of relativistic
degrees of freedom $N_{\rm eff} = 3.04$.

\subsection{$w$CDM fits to WiggleZ measurements}

As an initial analysis we fitted a flat $w$CDM cosmological model to
these datasets in which spatial curvature is fixed at $\Omega_{\rm k}
= 0$ but the equation-of-state $w$ of dark energy is varied as a free
parameter.  We fitted for the three parameters $(\Omega_{\rm m},
\Omega_{\rm m} h^2, w)$ using flat, wide priors which extend well
beyond the regions of high likelihood and have no effect on the
cosmological fits.  We only use the joint WiggleZ measurements of
$A(z)$ and $F(z)$ in these fits, not the growth rate data.  The extra
complexity in the normalization of the clustering pattern required to
fit $f \, \sigma_8(z)$ is analyzed by Parkinson et al.\ (in prep.).

\begin{figure}
\begin{center}
\resizebox{8cm}{!}{\rotatebox{270}{\includegraphics{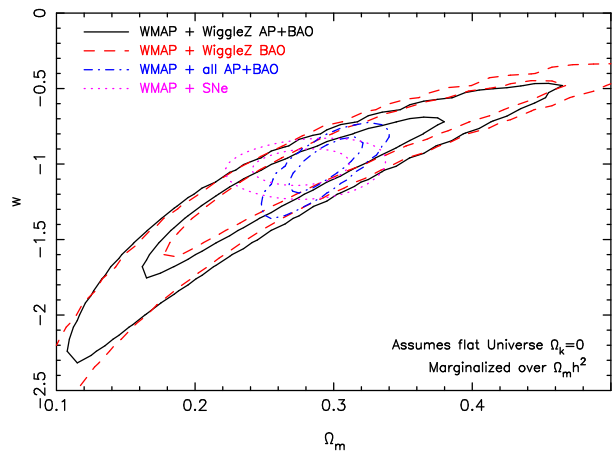}}}
\end{center}
\caption{The joint probability for parameters $\Omega_{\rm m}$ and $w$
  fitted to different datasets added to the WMAP distance priors,
  marginalizing over $\Omega_{\rm m} h^2$ and assuming $\Omega_{\rm k}
  = 0$.  The CMB data is combined in turn with the WiggleZ acoustic
  scale and Alcock-Paczynski distortion measurements (with appropriate
  covariance; black solid contours), the WiggleZ acoustic scale
  measurements alone (red dashed contours), all BAO measurements (blue
  dash-dotted contours), and SNe distance data (magenta dotted
  contours).  The two contour levels in each case enclose regions
  containing $68.27\%$ and $95.45\%$ of the total likelihood.}
\label{figprobomw}
\end{figure}

Figure \ref{figprobomw} displays the joint likelihood of the
parameters $(\Omega_{\rm m},w)$ marginalizing over $\Omega_{\rm m}
h^2$, comparing the effects of adding different datasets to the WMAP
distance priors.  The combination with the joint WiggleZ measurements
of $A(z)$ and $F(z)$ is illustrated by the (black) solid contours,
with the (red) dashed contours showing the improvement compared to
only using the WiggleZ measurements of $A(z)$.  The results are
consistent with, albeit with a significantly lower accuracy than,
parameter measurements based on the combination of the WMAP distance
priors with all WiggleZ+BOSS AP+BAO data, and all SNe data, which are
represented by the (blue) dash-dotted contours and (magenta) dotted
contours, respectively.

\subsection{$H(z)$ fits in bins}

\begin{figure*}
\begin{center}
\resizebox{16cm}{!}{\rotatebox{270}{\includegraphics{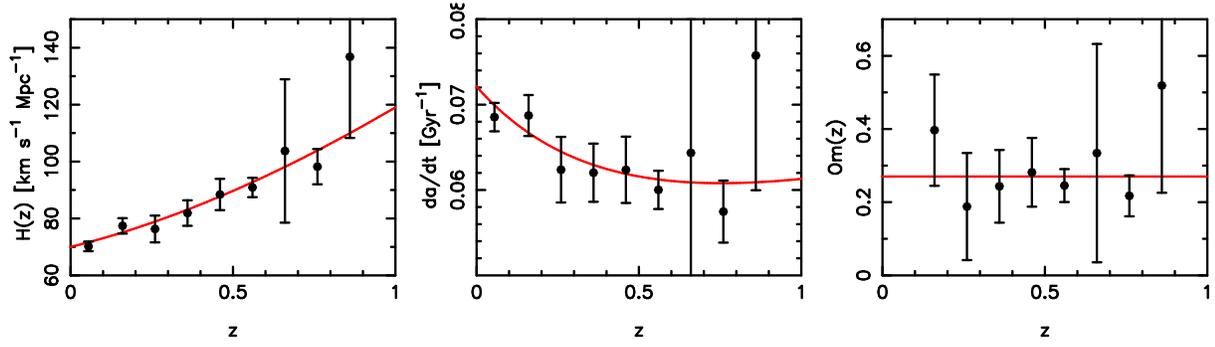}}}
\end{center}
\caption{Panels displaying, from left to right, measurements of the
  Hubble parameter $H(z)$, the cosmic expansion rate $\dot{a} =
  H(z)/(1+z)$, and the ``Om'' statistic $\left[ [H(z)/H_0]^2 - 1
    \right] / \left[ (1+z)^3 - 1 \right]$ fit as a stepwise function
  in 9 redshift bins of width $\Delta z = 0.1$ using a Monte Carlo
  Markov Chain.  The fit is performed to the WiggleZ and BOSS joint
  BAO and AP datasets, other BAO measurements from 6dFGS and SDSS, and
  SNe distance data.  The measurement in each bin is marginalized over
  $H(z)$ in the other bins, the spatial curvature $\Omega_{\rm k}$ and
  a WMAP prior for the sound horizon at baryon drag.  The solid lines
  are not fits to the data but represent a fiducial flat $\Lambda$CDM
  cosmological model with parameters $\Omega_{\rm m} = 0.27$ and $h =
  0.71$.  We do not plot a value for ${\rm Om}(z)$ in the first
  redshift bin because the statistic is not well-defined in the limit
  $z \rightarrow 0$.}
\label{fighzall}
\end{figure*}

We performed a model-independent determination of the cosmic expansion
history by carrying out a Monte Carlo Markov Chain (MCMC) fit for the
Hubble parameter $H(z)$ as a stepwise function in narrow redshift
bins, where we identify the value in the first bin as $H(z_1) = H_0 =
100 \, h$ km s$^{-1}$ Mpc$^{-1}$.  We also included the spatial
curvature $\Omega_{\rm k}$ as a free parameter in our fit, such that
we derived the angular diameter distance $D_A(z)$ from the co-moving
radial co-ordinate $r(z)$ as:
\begin{eqnarray}
D_A(z) = r(z)/(1+z) \hspace{0.5cm} &\Omega_{\rm k} = 0& \nonumber \\
D_A(z) = R_{\rm curv} \sinh{(r(z)/R_{\rm curv})}/(1+z) \hspace{0.5cm} &\Omega_{\rm k} > 0& \nonumber \\
D_A(z) = R_{\rm curv} \sin{(r(z)/R_{\rm curv})}/(1+z) \hspace{0.5cm} &\Omega_{\rm k} < 0&
\end{eqnarray}
where $R_{\rm curv} = c/(H_0 \sqrt{|\Omega_{\rm k}|})$.  If the range
of the $i$th redshift bin is $z_{i,{\rm min}} < z < z_{i,{\rm max}}$,
and the redshift at which we are evaluating a distance lies in the
$n$th bin, the co-moving radial co-ordinate $r(z)$ is deduced from the
stepwise $H(z)$ function as
\begin{eqnarray}
r(z) &=& \int_0^z \frac{c}{H(z')} \, dz' \nonumber \\ &=&
\sum_{i=1}^{n-1} \frac{c \, (z_{i,{\rm max}} - z_{i,{\rm
      min}})}{H(z_i)} + \frac{c \, (z - z_{n,{\rm min}})}{H(z_n)} .
\end{eqnarray}
Expressing this relation as a linear interpolation rather than a
stepwise function makes little difference to the results.  The
availability of the WiggleZ and BOSS Alcock-Paczynski distortion
measurements, with their direct sensitivity to $H(z)$, brings two
significant benefits to this analysis: a more precise determination of
$H(z)$ in stepwise bins, and a lower covariance between the
measurements in different bins.

The left-hand panel of Figure \ref{fighzall} illustrates the
measurements of $H(z)$ in $N=9$ stepwise redshift bins of width
$\Delta z = 0.1$, where the likelihood is computed using the WiggleZ
and BOSS joint BAO and AP datasets, the other BAO measurements from
6dFGS and SDSS, and the SNe Union 2 dataset.  We note that we do not
use the WMAP distance priors in this fit because of the uncertainty in
extrapolating the expansion rate beyond our bins, in the redshift
range $z_{N,{\rm max}} < z < z_*$, in order to deduce the value of
$D_A(z_*)$ which is required to evaluate the quantities $\ell_A$ and
$\mathcal{R}$.  However, we do marginalize over a WMAP-motivated
Gaussian prior in $\Omega_{\rm m} h^2$ with mean $0.1345$ and width
$0.0055$.  This prior is used when fitting to the BAO dataset, both
when using the acoustic parameter $A(z) \propto \sqrt{\Omega_{\rm m}
  h^2}$ and when calibrating the sound horizon scale $r_s(z_d)$.  In
the latter case we assume a fiducial baryon density $\Omega_{\rm b}
h^2 = 0.0226$.  We also marginalize over a wide uniform prior in
spatial curvature $-1 < \Omega_{\rm k} < 1$.

\begin{figure}
\begin{center}
\resizebox{8cm}{!}{\rotatebox{270}{\includegraphics{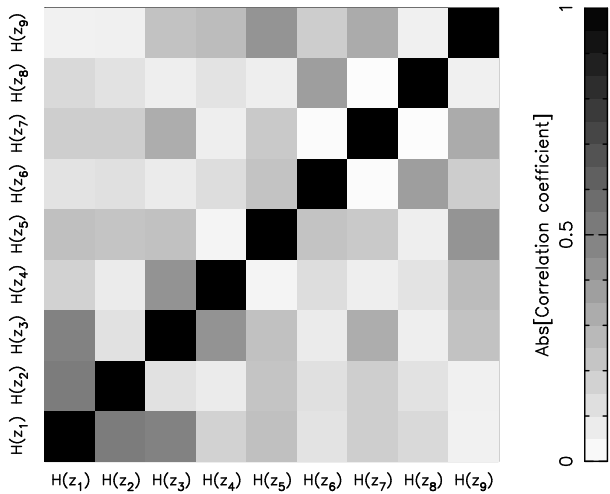}}}
\end{center}
\caption{The matrix of correlation coefficients for the measurements
  of the Hubble parameter $H(z)$ in 9 redshift bins plotted in Figure
  \ref{fighzall}, obtained using a Monte Carlo Markov Chain.}
\label{fighzcovall}
\end{figure}

We obtain measurements of $H(z)$ with precision better than $7\%$ in
most $\Delta z = 0.1$ redshift bins in the range $z < 0.8$ (we note
that the improved accuracy in the $0.7 < z < 0.8$ bin compared to the
adjacent bins is due to the presence of the WiggleZ Alcock-Paczynski
data point at $z=0.73$).  As displayed in Figure \ref{fighzall}, our
measurements are consistent with a flat $\Lambda$CDM cosmological
model with parameters $\Omega_{\rm m} = 0.27$ and $h = 0.71$ (the
value of the chi-squared statistic calculated using the full
covariance matrix is $7.52$ for 9 degrees of freedom).  Figure
\ref{fighzcovall} displays the covariance between the measurements of
$H(z)$ in each bin, deduced from the MCMC chain.  The correlation
coefficients between different bins vary depending on whether AP data
is available, but are generally low or moderate, $r < 0.5$.

Table \ref{tabhz} lists the marginalized measurements of $H(z)$ in
each bin.  We also convert these measurements to values of the cosmic
expansion rate $\dot{a} = da/dt$ in physical units, where $a =
1/(1+z)$ is the cosmic scale factor, and values of the ``Om''
statistic (Sahni, Shafieloo \& Starobinsky 2008) which is defined by
\begin{equation}
{\rm Om}(z) \equiv \frac{[H(z)/H_0]^2 - 1}{(1+z)^3 - 1} .
\end{equation}
In a spatially-flat $\Lambda$CDM model this statistic is constant at
different redshifts and equal to today's value of the matter density
parameter $\Omega_{\rm m}$.  In universes with different curvature, or
containing dark energy with different properties to a cosmological
constant, ${\rm Om}(z)$ would evolve with redshift.  These
determinations of $\dot{a}$ and ${\rm Om}(z)$ are plotted as the
central and right-hand panels in Figure \ref{fighzall}, respectively.
We see a significant decrease in the value of $\dot{a}$ between
redshifts $z=0$ and $=0.7$, corresponding to accelerating cosmic
expansion.  For example, the low-redshift expansion rate $\dot{a} =
0.069 \pm 0.002$ Gyr$^{-1}$ has dropped to $\dot{a} = 0.060 \pm 0.002$
Gyr$^{-1}$ at $z=0.55$ and $\dot{a} = 0.058 \pm 0.004$ at $z=0.75$.
The measurements of ${\rm Om}(z)$ are consistent with a constant
$\approx 0.25$, as expected in a spatially-flat $\Lambda$CDM model.

\begin{table}
\begin{center}
\caption{Measurements of the Hubble parameter $H(z)$, cosmic expansion
  rate $\dot{a} = H(z)/(1+z)$ and ``Om'' statistic $\left[
    [H(z)/H_0]^2 - 1 \right] / \left[ (1+z)^3 - 1 \right]$ fit as a
  stepwise function in 9 redshift bins of width $\Delta z = 0.1$ using
  a Monte Carlo Markov Chain.  The fit is performed to the WiggleZ and
  BOSS joint BAO and AP datasets, other BAO measurements from 6dFGS
  and SDSS-DR7, and SNe distance data.  The measurement in each bin is
  marginalized over $H(z)$ in the other bins, the spatial curvature
  $\Omega_{\rm k}$ and a WMAP prior for the sound horizon at baryon
  drag.  We do not quote a value for ${\rm Om}(z)$ in the first
  redshift bin because the statistic is not well-defined in the limit
  $z \rightarrow 0$.}
\label{tabhz}
\begin{tabular}{cccc}
\hline
$z$ & $H(z)$ & $\dot{a}(z)$ & ${\rm Om}(z)$ \\
& [km s$^{-1}$ Mpc$^{-1}$] & [Gyr$^{-1}$] & \\
\hline
$0.05$ & $70.2 \pm 1.7$ & $0.0685 \pm 0.0017$ & - \\
$0.15$ & $77.4 \pm 2.7$ & $0.0687 \pm 0.0024$ & $0.34 \pm 0.15$ \\
$0.25$ & $76.3 \pm 4.7$ & $0.0624 \pm 0.0038$ & $0.15 \pm 0.14$ \\
$0.35$ & $81.9 \pm 4.5$ & $0.0620 \pm 0.0034$ & $0.22 \pm 0.10$ \\
$0.45$ & $88.4 \pm 5.5$ & $0.0624 \pm 0.0039$ & $0.26 \pm 0.09$ \\
$0.55$ & $90.9 \pm 3.4$ & $0.0600 \pm 0.0022$ & $0.23 \pm 0.04$ \\
$0.65$ & $103.7 \pm 25.2$ & $0.0643 \pm 0.0156$ & $0.32 \pm 0.29$ \\
$0.75$ & $98.2 \pm 6.2$ & $0.0575 \pm 0.0036$ & $0.21 \pm 0.05$ \\
$0.85$ & $136.8 \pm 28.5$ & $0.0758 \pm 0.0158$ & $0.50 \pm 0.28$ \\
\hline
\end{tabular}
\end{center}
\end{table}

\begin{figure}
\begin{center}
\resizebox{8cm}{!}{\rotatebox{270}{\includegraphics{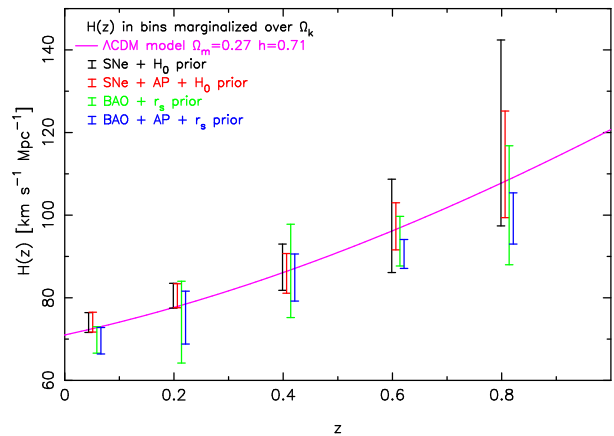}}}
\end{center}
\caption{Measurements of the Hubble parameter $H(z)$ fit as a stepwise
  function in 5 redshift bins of width $\Delta z = 0.2$ using a Monte
  Carlo Markov Chain.  Results are compared for datasets SNe + $H_0$
  prior, SNe + AP + $H_0$ prior, BAO + $r_s(z_d)$ prior, BAO + AP +
  $r_s(z_d)$ prior.  The measurement in each bin is marginalized over
  $H(z)$ in the other bins and the spatial curvature $\Omega_{\rm k}$.
  The solid line is not a fit to the data but represents a fiducial
  flat $\Lambda$CDM cosmological model with parameters $\Omega_{\rm m}
  = 0.27$ and $h = 0.71$.}
\label{fighzcases}
\end{figure}

In Figure \ref{fighzcases} we compare the results of fitting $H(z)$ in
5 bins of width $\Delta z = 0.2$ to different subsets of the total
dataset.  We note that, when combined with a CMB prior, the baryon
oscillation scale is calibrated in units of Mpc and permits direct
measurement of $H(z)$ in units of km s$^{-1}$ Mpc$^{-1}$.  However,
given that the normalization of the supernova Hubble diagram is
treated as an unknown parameter, the SNe data yield the relative
luminosity distance $D_L(z) H_0/c$ and require an extra prior in $H_0$
in order to determine the function $H(z)$.  We take this prior as the
Riess et al.\ (2011) $3\%$ determination of the Hubble constant using
new observations of Cepheid variables combined with Type Ia supernovae
and the megamaser host galaxy NGC 4258, which yields a Gaussian prior
in $H_0$ of mean $73.8$ km s$^{-1}$ Mpc$^{-1}$ and standard deviation
$2.4$ km s$^{-1}$ Mpc$^{-1}$.  Figure \ref{fighzcases} illustrates the
factor 2-3 gain in precision at higher redshifts achieved when adding
the WiggleZ and BOSS AP measurements, with their direct dependence on
$H(z)$, to the existing BAO and SNe dataset.  It is also notable that
the different probes produce consistent determinations of the
expansion history within the statistical errors, which are consistent
with a flat $\Lambda$CDM cosmological model with parameters
$\Omega_{\rm m} = 0.27$ and $h = 0.71$.

\begin{figure}
\begin{center}
\resizebox{8cm}{!}{\rotatebox{270}{\includegraphics{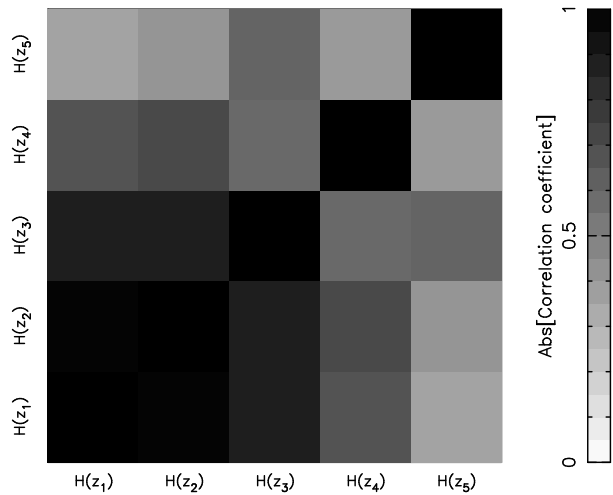}}}
\end{center}
\caption{The matrix of correlation coefficients for the measurements
  of the Hubble parameter $H(z)$ in 5 redshift bins using just the SNe
  dataset and $H_0$ prior, obtained using a Monte Carlo Markov Chain.}
\label{fighzcovsn}
\end{figure}

Another benefit of using the AP dataset is a reduced covariance
between the $H(z)$ measurements in different bins, which are heavily
correlated when only total distance information is available.  This is
illustrated by Figure \ref{fighzcovsn}, which displays the covariance
between the $H(z)$ measurements in 5 bins for the case of SNe data +
$H_0$ prior, illustrating the significantly higher correlation
coefficients $r > 0.5$ in comparison with Figure \ref{fighzcovall}.

In order to illustrate further the consistent results obtained when
varying the input datasets we fitted each determination of $H(z)$
plotted in Figure \ref{fighzcases} for the normalization factor $h$,
assuming $\Omega_{\rm m} = 0.27$, using the corresponding covariance
matrix obtained from the Markov chain.  The resulting measurements for
the cases (SNe, SNe+AP, BAO, BAO+AP) were $h = (0.72 \pm 0.03, 0.72
\pm 0.03, 0.69 \pm 0.02, 0.68 \pm 0.02)$, respectively.  [We note that
  the aim here is not to measure $h$, but to demonstrate that
  analyzing these subsets of the data produces consistent results.]

\subsection{Kinematical model fits}

Finally, we fitted our dataset with a ``kinematical'' cosmological
model (Rapetti et al.\ 2007) which is parameterized in terms of the
dimensionless second and third derivatives of the scale factor $a(t)$
with respect to time, the deceleration parameter $q(t) = -H^{-2}
(\ddot{a}/a)$ and jerk parameter $j(t) = H^{-3} (\dddot{a}/a)$.  In
particular, we adopt a parameterization where models are expressed in
terms of the present-day value of the deceleration parameter $q_0$ and
a constant jerk $j$, noting that $\Lambda$CDM models correspond to the
special case $j=1$.  Following Rapetti et al.\ (2007), for a given
$(q_0,j)$ we determine the function
\begin{equation}
V(a) = - \frac{\sqrt{a}}{2} \left[ \left( \frac{p-u}{2p} \right) a^p +
  \left( \frac{p+u}{2p} \right) a^{-p} \right] ,
\end{equation}
where $p = \frac{1}{2} \sqrt{1+8j}$ and $u = 2(q_0 + \frac{1}{4})$.
Given the function $V(a)$ we can determine the expansion rate as
\begin{equation}
\left[ \frac{H(z)}{H_0} \right]^2 = - \frac{2 \, V(a)}{a^2}
\end{equation}
We note that there is a region in the $(q_0,j)$ plane defined by
\begin{eqnarray}
j < q_0 + 2q_0^2 \hspace{1cm} &q_0& < -1/4 \nonumber \\
j < -1/8 \hspace{1cm} &q_0& > -1/4
\end{eqnarray}
for which the condition $V(a) \ge 0$ is not satisfied for all $a$ and
hence there is no Big Bang in the past.  We exclude this region from
our fits.

\begin{figure}
\begin{center}
\resizebox{8cm}{!}{\rotatebox{270}{\includegraphics{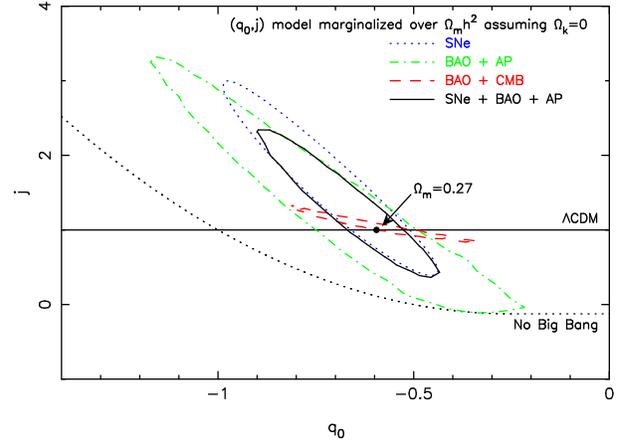}}}
\end{center}
\caption{The joint probability of kinematical model parameters $q_0$
  and $j$ fitted to different combinations of datasets and assuming
  $\Omega_{\rm k} = 0$.  Results are shown for SNe data alone, the BAO
  + AP dataset and BAO + WMAP.  The two contour levels in each case
  enclose regions containing $68.27\%$ and $95.45\%$ of the total
  likelihood.}
\label{figprobq0j}
\end{figure}

Figure \ref{figprobq0j} illustrates the joint likelihood of
kinematical model fits to the BAO, AP and SNe datasets.  As in
previous Sections, we marginalize over a WMAP-inspired Gaussian prior
in $\Omega_{\rm m} h^2$ in order to calibrate the baryon oscillation
standard ruler.  $\Lambda$CDM models correspond to the line $j=1$, and
specific values of $\Omega_{\rm m}$ pick out a point with $q_0 =
\frac{3}{2} \Omega_{\rm m} - 1$.  These models are consistent with the
data.  We note that SNe provide the best constraints on the
kinematical model parameters, and that including the WMAP distance
priors in the fitted dataset produces a very accurate joint constraint
on $(q_0,j)$ from the precisely-known distance to the last-scattering
surface, although this corresponds to a significant extrapolation of
the validity of the model from $a > 0.5$ to $a > 0.001$.  Fitting to
the combination of SNe, BAO and AP data, not including the CMB,
produces marginalized parameter measurements $q_0 = -0.67 \pm 0.16$
and $j = 1.37 \pm 0.68$.

\section{Conclusions}
\label{secconc}

We have used large-scale structure measurements from the WiggleZ Dark
Energy Survey to perform joint fits for the baryon-oscillation
distance scale quantified by the acoustic parameter $A(z) \propto
[D_A(z)^2/H(z)]^{1/3}$, the Alcock-Paczynski distortion parameter
$F(z) \propto D_A(z) H(z)$ and the normalized growth rate $f \,
\sigma_8(z)$ in three overlapping redshift slices with effective
redshifts $z = (0.44, 0.6, 0.73)$.  We use lognormal realizations to
quantify the covariances between parameters and redshift slices,
producing a $9 \times 9$ covariance matrix.

By combining the joint measurements of $A(z)$ and $F(z)$ taking into
account the covariance, we performed simultaneous determinations of
the angular-diameter distance $D_A(z)$ and Hubble parameter $H(z)$
based only on the WiggleZ Survey dataset and a WMAP prior in the
matter density $\Omega_{\rm m} h^2$ to calibrate the baryon
oscillation standard ruler.  These results are generally insensitive
to the fiducial cosmological model including spatial curvature.  We
measure these parameters with $7-9\%$ accuracy in each redshift bin.

We use a combined dataset consisting of these joint WiggleZ geometric
measurements, other BAO data and SNe luminosity distances to perform a
Monte Carlo Markov Chain determination of the expansion history $H(z)$
as a stepwise function in 9 redshift bins of width $\Delta z = 0.1$,
also marginalizing over spatial curvature $\Omega_{\rm k}$.  The
results are consistent with a flat $\Lambda$CDM cosmological model
with parameters $\Omega_{\rm m} = 0.27$ and $h = 0.71$.  The addition
of the AP data reduces both the errors in these measurements [through
  its direct sensitivity to $H(z)$] and the covariance between
different redshift bins.  When we convert our results to a measurement
of the cosmic expansion rate $\dot{a} = H(z)/(1+z)$, we see a
significant decrease in the value of $\dot{a}$ between redshifts $z=0$
and $=0.7$, corresponding to accelerating cosmic expansion.
Measurements of the statistic ${\rm Om}(z) = \left[ [H(z)/H_0]^2 - 1
  \right] / \left[ (1+z)^3 - 1 \right]$ are constant with redshift,
consistent with a spatially-flat $\Lambda$CDM model with matter
density parameter $\Omega_{\rm m} \approx 0.25$.

We compare our measurements to cosmological models including different
dark energy equations-of-state $w$ and kinematical models expressed in
terms of the derivatives of the cosmic scale factor, the deceleration
and jerk parameters.  We find all data to be consistent with a
cosmological constant model.

\section*{Acknowledgments}

We thank the anonymous referee for very useful feedback which greatly
improved the presentation of this paper.  CB acknowledges useful
discussions with Eric Linder, Berian James, Eiichiro Komatsu, David
Rapetti, Steve Allen, Eyal Kazin, Licia Verde and Raul Jimenez, and
thanks the astronomy groups at Berkeley and Stanford for hospitality
during the completion of this work.  We acknowledge financial support
from the Australian Research Council through Discovery Project grants
DP0772084 and DP1093738 and Linkage International travel grant
LX0881951.  SC and DC acknowledge the support of an Australian
Research Council QEII Fellowship.  We are also grateful for support
from the Centre for All-sky Astrophysics, an Australian Research
Council Centre of Excellence funded by grant CE11E0090.

\end{document}